\documentclass[aps,pre,twocolumn,superscriptaddress,showpacs]{revtex4}

\usepackage{graphicx}
\usepackage{amssymb}
\usepackage{epsfig}
\usepackage{amsmath}
\usepackage{times}
\usepackage{float}
\usepackage{epstopdf}
\usepackage{color}
\begin{document}

\title{Nontrivial standing wave state in frequency-weighted Kuramoto model}

\author{Hongjie Bi}\thanks{Authors contributed equally to this work.}
\affiliation{Department of Physics, East China Normal University, Shanghai 200241, China}

\author{Yan Li}\thanks{Authors contributed equally to this work.}
\affiliation{Nantong Middle School,
9 Zhongxuetang Road, Nantong, Jiangsu 226001, China }
\affiliation{Department of Physics, East China Normal University, Shanghai 200241, China}

\author{Li Zhou}\thanks{Corresponding author: smilingzl@163.com}
\affiliation{No. 4 Middle School Affiliated to ECNU, 279 Luding Road, Shanghai 200062, China}

\author{Shuguang Guan}\thanks{Corresponding author: guanshuguang@hotmail.com}
\affiliation{Department of Physics, East China Normal University, Shanghai 200241, China}

\date{\today}

\begin{abstract}

Synchronization in a frequency-weighted Kuramoto model with a uniform frequency distribution is studied.
We plot the bifurcation diagram and identify the asymptotic coherent states.
Numerical simulations show that the system undergoes two first-order transitions in both the forward and backward directions.
Apart from the trivial phase-locked state, a novel nonstationary coherent state, i.e., a nontrivial standing wave state is observed and characterized.
In this state, oscillators inside the coherent clusters are not frequency-locked as they would be in the usual standing wave state.
Instead, their average frequencies are locked to a constant.
The critical coupling strength from the incoherent state to the nontrivial standing wave state can be obtained by performing linear stability analysis. The theoretical results are supported by the numerical simulations.\\


\end{abstract}


\maketitle

\section{Introduction}

Synchronization phenomena are collective behaviors emerging in dynamical systems, which are widely observed in physics, chemistry, biology, and social science \cite{pikovsky2003synchronous,huang2007}.
Synchronization in systems of coupled oscillators has been extensively investigated theoretically. For example,
the Kuramoto model \cite{kuramoto1975} is a successful prototype in studies of synchronization.
The Kuramoto model and its variants have been studied for decades and have been found to exhibit a continuous (second-order) phase transition from incoherence to synchrony \cite{strogatz2000,crawford1994}.
However, recent works reveal that
a discontinuous (first-order) synchronization transition could also occur in some generalized Kuramoto models \cite{gomez2011,zhang2013,hu2014,zou2014,zhou2015,zhang2015};
for example, the
Kuramoto model in a scale-free (SF) network \cite{gomez2011,xu2016} and a star network \cite{zou2014},
the frequency-weighted Kuramoto model \cite{zhang2013,hu2014,zhou2015},
the Kuramoto model in adaptive and multilayer networks \cite{zhang2015}, and the Kuramoto model with both conformists and contrarians \cite{hong2011prl}.

In particular, the frequency-weighted Kuramoto model has been intensively investigated under various frequency distributions,
including unimodal distributions (Lorentzian, Gaussian, and triangular) \cite{zhang2013,hu2014}, the half-Gaussian distribution \cite{zhang2013}, asymmetric unimodal distributions (Lorentzian, Gaussian, triangular, and Rayleigh) \cite{zhou2015}, and the bimodal distribution \cite{hu2014,bi2016}.
Nevertheless, for the most simple frequency distribution, i.e., the uniform distribution, a thorough study of both the bifurcation diagram and the possible coherent states is still lacking.
For this purpose, in this work, we focus on the first-order synchronization (i.e., explosive synchronization) transitions in a frequency-weighted Kuramoto model with a uniform frequency distribution.

We obtain two main results via both theoretical and numerical studies. First, a detailed bifurcation diagram is plotted. Interestingly, it shows that as the coupling strength increases/decreases, the system successively undergoes two first-order transitions toward synchronization/incoherence, which are typically characterized by hysteresis loops when both forward and backward transitions are considered.
Second, we identify three asymptotic states to which the long-term dynamics of the system evolves, i.e., the incoherent state, the nontrivial standing wave (NSW) state (see detailed description below), and the phase-locked state.
Remarkably, in the NSW state, the synchronous oscillators split into two clusters, which are always counter-rotating, just as in the usual standing wave state. However, in these coherent clusters, the instantaneous frequencies of the oscillators are not the same.
Instead, their average frequencies are equal to constant values.
This NSW state, in fact, is a type of nonstationary coherent state in a coupled oscillator system.
Here, ``nonstationary'' means that in this state, the probability density function of the oscillators is time-dependent.
Through linear stability analysis, we analytically obtain the critical coupling strength at which the system bifurcates from the incoherent state to the NSW state. In addition, the bifurcation boundaries between the three asymptotic states are identified numerically.

This paper is organized as follows. In Sec. II, we briefly introduce the frequency-weighted Kuramoto model with a uniform frequency distribution. In Sec. III, we discuss the bifurcation diagram and characterize the NSW state in detail. Sec. IV discusses the linear stability of the incoherent state. Finally, we summarize our work in Sec. V.

\begin{figure*}[!htbp]
  \centering
  \includegraphics[width=0.95\textwidth]{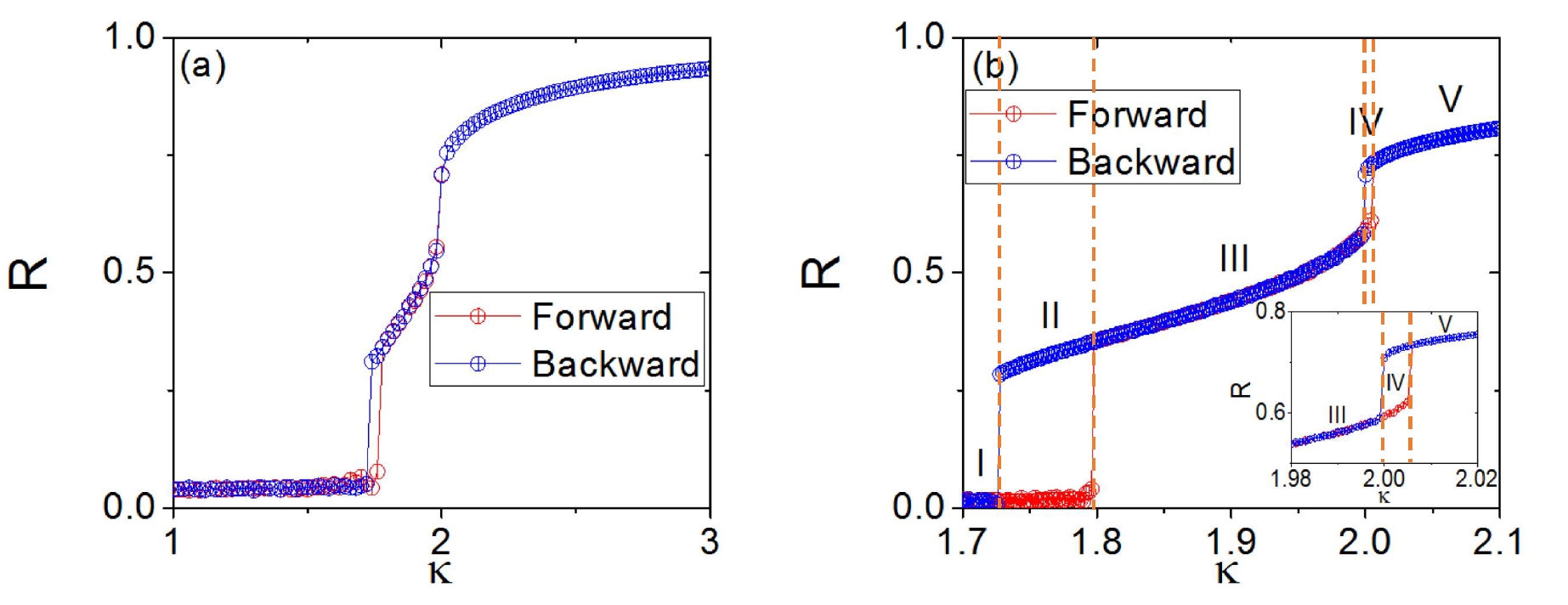}
  \caption{(color online) (a) Bifurcation diagram ($R$ vs. $\kappa$) characterizing both the forward and backward synchronization transitions in model (\ref{eq:model}) with a uniform frequency distribution. (b) Enlargement of (a) that clearly shows two hysteresis loops, i.e., two first-order transitions, and five dynamical regimes.
  As $\kappa$ increases, the first forward transition occurs at $\kappa_{f1}=1.798$, where the system jumps from the incoherent state to the NSW state. Then, at the second transition point, $\kappa_{f2}=2.006$, the NSW state jumps into the phase-locked state. For the backward direction, the system first jumps from the phase-locked state to the NSW state at $\kappa_{b2}=2.00$. Then the NSW state jumps into the incoherent state at $\kappa_{b1}=1.728$.
  $N=10,000, \gamma=0.5$.}\label{fig1}
\end{figure*}

\section{The dynamical model}

In the frequency-weighted Kuramoto model \cite{zhang2013,hu2014,zhou2015}, $N$ phase oscillators are governed by the following dynamical equation:
\begin{equation}\label{eq:model}
\dot{\theta}_{i}=\omega_{i}+\frac{\kappa|\omega_{i}|}{N}\sum^{N}_{j=1}\sin(\theta_{j}-\theta_{i}),~~~~~i=1,\cdots\cdots N,
\end{equation}
where $\theta_i$ and $\omega_{i}$ are the phase and natural frequency of the $i$th oscillator, respectively. The dot denotes a time derivative, and $\kappa$ is the global coupling strength.
The most important characteristic of this model is that the $i$th oscillator is coupled to the mean field via the effective coupling strength $\kappa|\omega_{i}|$, which is proportional to the magnitude of its natural frequency. In the mean-field approximation, this model is equivalent to the classical
Kuramoto model in a SF network \cite{gomez2011}.

Typically, the natural frequencies of oscillators in Kuramoto-like models are drawn from a certain distribution $g(\omega)$. As mentioned above, this model has been studied under various frequency distributions other than the simple uniform distribution to date. Therefore, in this work, we investigate this system when the natural frequencies satisfy the following uniform distribution:
\begin{equation}\label{eq:frequency}
g(\omega)=
\begin{cases}
\frac{1}{2\gamma} & \textrm{for} ~ |\omega|\leq \gamma,\\
0 & \textrm{for} ~ |\omega|>\gamma.
\end{cases}
\end{equation}
Without loss of generality, we take $\gamma=\frac{1}{2}$ in the frequency distribution in this work.

In order to characterize the degree of phase coherence in the model, an order parameter can be defined as
\begin{equation}\label{eq:order}
R e^{i\psi}=\frac{1}{N}\sum_{j=1}^{N}e^{i\theta_j},
\end{equation}
where $R$ and $\psi$ are the module and argument of the mean field, respectively. Geometrically, the complex order parameter can be regarded as a vector on the complex plane.
By definition, $R$ is between 0 and 1. Typically, $R=0$ indicates a totally random phase distribution, i.e., the incoherent state, whereas $R>0$ indicates a (partially) phase-locked state, i.e., the coherent or synchronized state. As the system becomes more coherent, $R$ will gradually approach 1.

In this work, coupled ordinary differential equations are numerically integrated using the
fourth-order Runge--Kutta method with a time step of 0.01. The initial conditions for the phase variables are random.
Typically, the total number of oscillators is $N$ = 10,000.
We explore both the forward and backward
transitions adiabatically to test whether hysteresis exists in the synchronization transitions. For each control parameter, the order parameter is averaged in a time window after
the transient stage. Such numerical schemes are adopted throughout this paper.

\begin{figure*}[!htbp]
  \centering
  \includegraphics[width=1.0\textwidth]{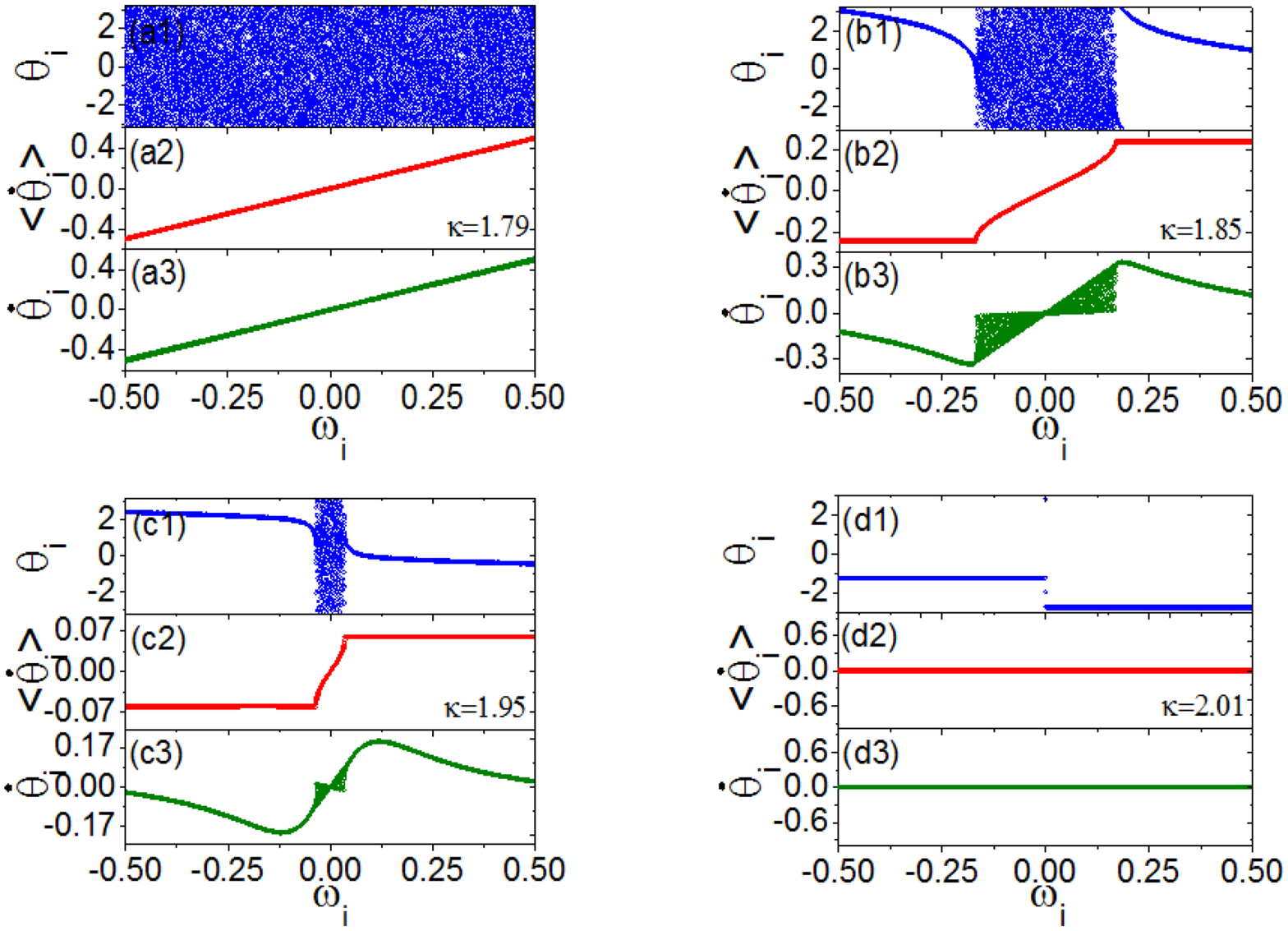}
  \caption{(color online)
Typical asymptotic states in model (\ref{eq:model}) with a uniform frequency distribution during the forward transition.
Snapshots of the instantaneous phases $\theta_i$ (upper plots), the average frequencies (average speeds) $\langle\dot{\theta_i}\rangle$ (middle plots), and the instantaneous frequencies (speeds) $\dot{\theta_i}$ (lower plots) vs. natural frequencies $\{\omega_i\}$ of the oscillators. (a) Incoherent state at $\kappa=1.79$. (b) NSW state at $\kappa=1.85$. (c) NSW state at $\kappa=1.95$. (d) Phase-locked state at $\kappa=2.01$.
As shown in Fig. 1,
the system bifurcates first from the incoherent state to the NSW state and then to the phase-locked state, and these two transitions are both first-order ones.
In (b), two coherent clusters are formed, where their average frequencies are distributed symmetrically with respect to 0, i.e., the cluster with positive frequencies and that with negative frequencies have the same average rotational speed but rotate in opposite directions (b2). However, the instantaneous frequencies (speeds) of oscillators in the coherent clusters are time-dependent and generally differ from one another (b3).
In (c), as the coupling strength increases, more oscillators are entrained to join the coherent clusters.
In the phase-locked state (d), the instantaneous frequencies of the coherent oscillators are locked to a constant. Therefore, this state is stationary and it differs essentially from the nonstationary NSW states in (b) and (c).
}\label{fig2}
\end{figure*}

\section{Bifurcations and the NSW state}

We first study the bifurcation diagram of model (\ref{eq:model}) with a uniform frequency distribution. The results of extensive numerical simulations are summarized in Fig. 1.
Specifically, we observed three types of asymptotic states: the incoherent state, NSW state, and phase-locked state. Interestingly, as the coupling strength increases, we find two first-order phase transitions [Fig. 1]. The first is from the incoherent state to the NSW state, and the second is from the NSW state to the phase-locked state. Inversely, when the system starts from the coherent state and as the coupling strength decreases, it also experiences two first-order phase transitions. However, the bifurcation points are smaller than those of the forward transitions. We use $\kappa_{f1}$, $\kappa_{f2}$, $\kappa_{b1}$, and $\kappa_{b2}$ to denote the critical points. Here the subscripts $f1$ and $b1$ denote the forward and backward critical points, respectively, for the first transition, and $f2$ and $b2$ are those for the second transition.
Numerically, they are identified as
$\kappa_{f1}=1.798$, $\kappa_{f2}=2.006$, $\kappa_{b1}=1.728$, and $\kappa_{b2}=2.00$.
Thus,
the bifurcation diagram can be divided into five parameter regimes
as follows:
(I) $0 < \kappa < \kappa_{b1}$, where only the incoherent state is stable;
(II) $\kappa_{b1} < \kappa < \kappa_{f1}$, where both the incoherent state and the NSW state are stable;
(III) $\kappa_{f1} < \kappa < \kappa_{b2}$, where only the NSW state is stable;
(IV) $\kappa_{b2} < \kappa < \kappa_{f2}$, where both the NSW state and the phase-locked state are stable; and
(V) $\kappa_{f2} < \kappa$, where only the phase-locked state is stable.
These five parameter regimes are clearly shown in Fig. 1(b).

We now investigate the asymptotic states in model (\ref{eq:model}) with a uniform frequency distribution.
The long-term dynamics of the system are found to evolve to one of three states: the incoherent state, NSW state, or phase-locked state. In the following, we take the forward transition process as an example for description. When the coupling strength is below the critical point $\kappa_{f1}$, all the oscillators are desynchronized. This incoherent state is shown in Fig. 2(a). When $\kappa_{f1}<\kappa<\kappa_{f2}$, the system evolves to a new NSW state. Figs. 2(b) and (c) demonstrate two such examples at $\kappa=1.85$ and $\kappa=1.95$, respectively. In these states, the oscillators in the system have been only partially entrained. The coherent oscillators split into two groups corresponding to the positive and negative natural frequencies, respectively. They coexist with the
drifting (desynchronized) ones. These two coherent clusters counter-rotate along the unit circle. Overall, this scenario resembles the normal SW state \cite{crawford1994,martens2009}.
Interestingly, however, this NSW state differs essentially from
the normal SW state, in which the oscillators within each cluster are frequency-locked. A careful examination of the NSW state reveals that the oscillators in the coherent clusters are not frequency-locked; i.e., their instantaneous frequencies are not the same.
This novel property can be clearly seen in Figs. 2(b3) and (c3), where the snapshots show that the instantaneous frequencies inside either coherent cluster are not locked.
However, they are correlated in a certain way such that, surprisingly, the average frequencies are locked to a constant. These important characteristics can be immediately seen by comparing the smooth cusped patterns in Figs. 2(b3) and (c3) with the staircase structures in Figs. 2(b2) and (c2).

\begin{figure*}[!htpb]
  \centering
  \includegraphics[width=1.0\textwidth]{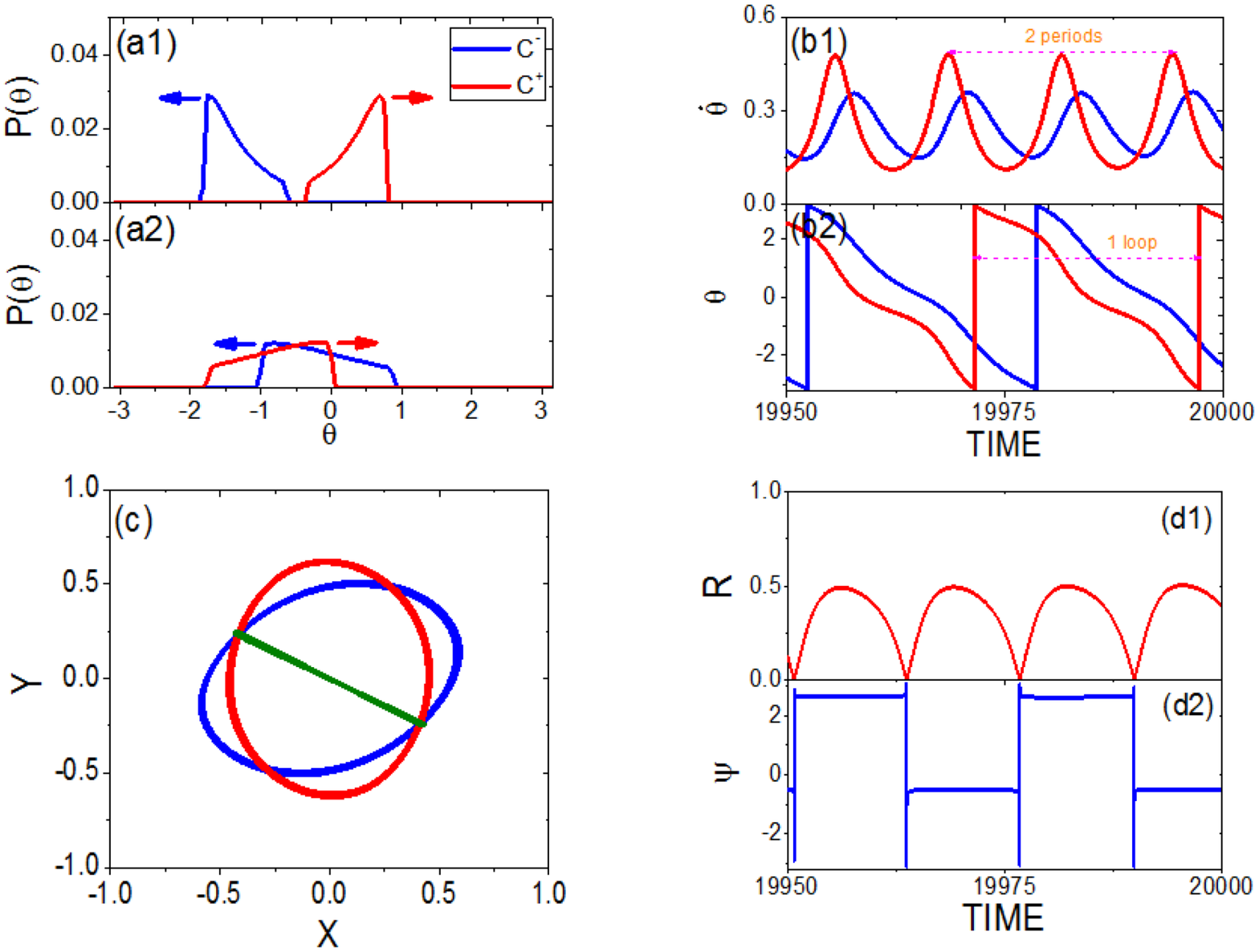}
   \caption{(color online)
   Characterization of the NSW state in Fig. 2(b).
    (a) Snapshots of the instantaneous phase distributions at different moments in one period. Red and blue denote clusters with positive and negative frequencies, respectively.
    In this nonstationary state, two coherent clusters rotate in opposite directions with nonuniform speeds, and the shapes of the clusters are also time-dependent.
    (b) Time series of the instantaneous speeds (b1) and instantaneous phases (b2) for two sample coherent oscillators.
    Although the instantaneous frequencies of these two oscillators are different, their average frequencies during one period are the same. A comparison of (b1) with (b2) reveals that during one loop along the unit circle, the instantaneous speeds of the oscillators exhibit two periods.
   (c) Order parameters for all oscillators (including the drifting ones) with positive (red oval) and negative (blue oval) frequencies, and the order parameter for all oscillators (green line).
   (d) Time series of the global order parameters $R(t)$ and $\psi(t)$, which are typically oscillatory.
   As the global order parameter oscillates approximately along the green line in (c), its phase is found to be binary, as shown in (d2).    }
  \label{fig3}
 \end{figure*}

In Fig. 3, we further characterize the typical NSW state corresponding to Fig. 2(b) from both macroscopic and microscopic perspectives.
As the instantaneous frequency characterizes the rotations of the oscillators along the unit circle, the system exhibits a very interesting collective motion of the oscillators.
Fig. 3(a) shows two snapshots of the instantaneous phase distributions. The system clearly splits into two symmetric groups that are counter-rotating. Furthermore, the shapes of the instantaneous phase distributions change continuously within one period. This implies that
the oscillators inside one coherent cluster are correlated in a complicated way, rather than simply being frequency-locked. In the latter case, i.e., the usual SW state, the oscillators inside one coherent cluster behave like a giant cluster, and the phase distribution does not change its shape but keeps moving as a whole.
In Fig. 3(b1), we show that the instantaneous frequencies of oscillators inside the same cluster evolve periodically, but different oscillators follow different periodic patterns. In other words, the instantaneous rotational speed of each oscillator varies uniquely with time. This characteristic can also be seen in Fig. 3(b2), where the evolutions of the instantaneous phases corresponding to Fig. 3(b1) are plotted.
This novel feature makes the observed NSW state essentially different from the usual SW states previously reported in Kuramoto-like models, in which oscillators are typically frequency-locked inside the coherent cluster \cite{crawford1994,martens2009}.
In brief, it is the average frequencies rather than the instantaneous frequencies that are locked in this NSW state. The order parameters for the coherent clusters collectively exhibit complicated orbits in phase space, as shown in Fig. 3(c). The amplitude of the overall order parameter oscillates nearly periodically with time [Fig. 3(d1)]. In addition, its phase is found to be approximately binary during its evolution [Fig. 3(d2)].

We emphasize that during the backward transition, the NSW state can also be observed when $\kappa_{b1}<\kappa<\kappa_{b2}$. As the coupling strength decreases, the system first jumps into the NSW state from the coherent state and then jumps into the incoherent state when $\kappa$ is below $\kappa_{b1}$.

Note that the above results are based on numerical studies of a specific system with size $N=10,000$. In fact, we have conducted extensive simulations with different system sizes, for example, $N=5000$ and $N=20,000$. We found that the size effect could change the bifurcation points slightly, particularly the forward ones. Nevertheless, the entire bifurcation diagram is qualitatively the same.

\section{Critical coupling strength}

From a theoretical viewpoint, it is desirable to solve the critical coupling strength for the system in the thermodynamic limit.
Recently, a method based on the Ott--Antonsen (OA) ansatz was successfully used for this purpose in many systems \cite{Ott2008}. It can effectively reduce the dynamics of coupled oscillators to a low-dimensional manifold and thus, greatly facilitate the analysis. However, the use of the OA method depends on several conditions. In particular, it requires
analytical continuation. In the frequency-weighted model, i.e., Eq. (\ref{eq:model}), there is a term of absolute value $|\omega_i|$, which hinders analytical continuation when the OA method is applied.
, the OA method has failed to treat this model to date. In the following, we turn to the traditional method of linear stability analysis.

In order to obtain the critical coupling strength $\kappa_{f1}$ for the synchronization transition in model (\ref{eq:model}), we perform linear stability analysis of the incoherent state.
In the mean-field form, Eq. (\ref{eq:model}) can be written as
\begin{equation}\label{eq:rmodel}
\dot{\theta}_{i}=\omega_{i}+\kappa|\omega_{i}|R\sin(\psi-\theta_{i}),
\end{equation}
where $R$ and $\psi$ are the order parameters defined in Eq. (\ref{eq:order}).
Following the analysis in Refs. \cite{strogatz2000,hu2014}, the critical equation relating the coupling strength $\kappa$ and the eigenvalue $\lambda$ is \cite{hu2014}
\begin{equation}\label{eq:critical}
\frac{2}{\kappa}=\int_{-\infty}^{+\infty}\frac{\lambda|\omega|}{\lambda^{2}+\omega^{2}}g(\omega)d\omega.
\end{equation}
Substituting the uniform distribution of Eq. (2) into the integration of Eq. (5), we get
\begin{equation}\label{eq:integration}
\frac{2}{\kappa}=2\int_{0}^{\gamma}\frac{\lambda\omega}{\lambda^{2}+\omega^{2}}\cdot\frac{1}{2\gamma}d\omega=\frac{\lambda}{2\gamma}\ln(1+\frac{\gamma^{2}}{\lambda^{2}}).
\end{equation}
Then Eq. (5) becomes
\begin{equation}\label{eq:integration1}
\frac{4\gamma}{\kappa}=\lambda\ln(1+\frac{\gamma^{2}}{\lambda^{2}})=f(\lambda).
\end{equation}
When $\lambda$ crosses the imaginary axis, the incoherent state
loses its stability.
Thus, setting $\lambda=iy$, we have
\begin{equation}\label{eq:integration2}
 f(iy)=iy\ln(1-\frac{\gamma^{2}}{y^{2}}).
\end{equation}
If $|y|>\gamma$, $f(iy)$ is a pure imaginary number. Thus, Eq. (7) has no solution because its left-hand side is real.
If $|y|<\gamma$, $1-\frac{\gamma^{2}}{y^{2}}<0$, and $\ln(1-\frac{\gamma^{2}}{y^{2}})=\ln(\frac{\gamma^{2}}{y^{2}}-1)-i\pi$.
We have
\begin{equation}\label{eq:integration3}
f(iy)=iy\ln(\frac{\gamma^{2}}{y^{2}}-1)+\pi y.
\end{equation}
Combining Eq. (7) with Eq. (9), we obtain
\begin{eqnarray}\label{eq:result}
y\ln(\frac{\gamma^{2}}{y^{2}}-1) &=& 0,\\
\frac{4\gamma}{\kappa_{f1}} &=& \pi y.
\end{eqnarray}
From Eq. (10), $y$ has a nontrivial solution $y=\frac{\gamma}{\sqrt{2}}$, which leads to the critical coupling strength
\begin{equation}\label{eq:integration3}
\kappa_{f1}=\frac{4\sqrt{2}}{\pi}\approx 1.8.
\end{equation}
The above analysis reveals that
$\kappa_{f1}$ is independent of the width of the uniform distribution.
Physically, this arises from the properties of the frequency-weighted model.
As shown in Eq. (4), the effective coupling strength for an oscillator interacting with the mean field is proportional to its frequency. Thus, the influence of the natural frequency is balanced by the effective coupling strength.

Moreover, according to Ref. \cite{hu2014},
the critical coupling strength $\kappa_{b2}$ is universal, regardless of the details of specific frequency distributions. For a symmetric unimodal frequency distribution,
\begin{equation}\label{eq:integration6}
\kappa_{b2}=2
\end{equation}
still holds. During the backward transition, the system bifurcates from the phase-locked state to the NSW state at this point.

Note that we cannot yet provide a suitable description of the NSW state. Therefore, it is difficult to obtain the other two bifurcation points, $\kappa_{f2}$ and $\kappa_{b1}$, by applying linear stability analysis to the NSW state.

\bigskip
\section{Conclusion}

We studied explosive synchronization in a frequency-weighted Kuramoto model with a uniform frequency distribution.
Numerical simulations and theoretical analysis revealed that the system exhibits two first-order transitions in both the forward and backward directions. In these transitions, the system bifurcates among three asymptotic states: the incoherent state, NSW state, and phase-locked state.
Among them, the NSW state, a new coherent state in this model, was further characterized. In the coherent clusters in this state, it is the average frequencies rather than the instantaneous frequencies of the oscillators that are locked. Physically, the NSW state is a weaker coherence (compared with the phase-locked state) achieved by the coupled oscillators when the coupling strength is in the intermediate regime.
It is between the incoherent state (full asynchrony) and the phase-locked state (full synchrony) and thus exhibits remarkable dynamical properties.
Recently, a quantized, time-dependent clustered state,
called the Bellerophon state,
was revealed in globally coupled oscillators \cite{bi2016,Qiu2016}.
The NSW state reported in this paper can be classified as a special type of Bellerophon state in which only two coherent clusters coexist.
These results are helpful for understanding the complicated collective behavior in generalized Kuramoto models.

\begin{acknowledgments}

This work is supported by the National Natural
Science Foundation of China under Grant
No. 11135001.

\end{acknowledgments}

\end{document}